\documentclass{ws-ijbc}
\usepackage{dcolumn}
\usepackage{bm}
\usepackage{changebar,color,amsmath,rotating,eso-pic,booktabs,tabularx,wrapfig,fontenc,natbib}

\usepackage{multirow}

\begin{document}

\title{Quantifying Changes in the Spatial Structure of Trabecular Bone}
\author{Norbert Marwan$^{1,2,}$\thanks{email: marwan@pik-potsdam.de},
Gise Beller$^{3}$,
Dieter Felsenberg$^{3}$,\\
Peter Saparin$^{4,}\thanks{Presently with Perceptive Informatics, a PAREXEL technology company, Am Bahnhof Westend 15, 14059 Berlin, Germany}$,
J\"urgen Kurths$^{1,5}$\\
\small{$^1$ Potsdam Institute for Climate Impact Research (PIK)}\\
\small{14412 Potsdam, Germany}\\
\small{$^2$ Interdisciplinary Center for Dynamics of Complex Systems, University of Potsdam}\\
\small{14415 Potsdam, Germany}\\
\small{$^3$ Charit\'e, Campus Benjamin Franklin, University of Medicine Berlin}\\
\small{12200 Berlin, Germany}\\
\small{$^4$ Department of Biomaterials, Max Planck Institute of Colloids and Interfaces}\\
\small{14424 Potsdam-Golm, Germany}\\
\small{$^5$ Department of Physics, Humboldt University Berlin}\\
\small{12489 Berlin, Germany}\\
}

\date{\today}

\maketitle

\begin{abstract}
We apply recently introduced measures of complexity for the structural
quantfication of distal tibial bone. For the first time, we are able
to investigate the temporal structural alteration of trabecular bone.
Based on four patients, we show how bone may alter due to temporal
immobilisation.\\


\end{abstract}

\clearpage

\section{Introduction}
The whole human body is a highly complex and dynamic
system ensuring adaptation to a changing environment.
Consequently, the human skeletal system is continuously
changing in order to adapt its strength especially to
current loading conditions. However, in subjects suffering on
certain bone diseases, like osteoporosis (loss of bone
mineral density and increased) fracture risk), limited moveability, or 
staying in micro-gravity conditions, 
bone may change so dramatic that the bone will lose a 
significant amount of its stability and fracture risk 
increases. These changes are on the one hand loss of bone mass,
a decrease of the mineralisation of bone, and on the other
hand a change in the micro-architecture of the interior bone
(trabecular bone). Structural changes in trabecular bone have 
received much attention in the last years because the loss of bone mass
alone cannot explain all variation in bone strength. Moreover,
the rapid progress in high resolution Computed Tomography 
(CT) imaging facilitates
the investigation of the micro-architecture of bone. 

The standard method for the quality assessment of the micro-architecture 
of trabecular bone is histomorphometry \cite{parfitt1983,ito1998, hildebrand1999}. 
More recent and advanced approaches quantify the complexity of 
trabecular structures by using measures of complexity based on symbolic dynamics 
\cite{saparin1998,saparin2005}, recurrence \cite{marwan2007pla}, fractal 
\cite{marwan2007epjst} or geometric properties \cite{marwan2009}, or using
volumetric spatial decompositions \cite{stauber2006c}. 
Applying these approaches on 3D images of trabecular bone, it was
shown that the micro-architecture changes substantially
during the development 
of osteoporosis. The main conclusions in 
\cite{saparin2005,marwan2007epjst,marwan2009}
were that the complexity of the bone micro-architecture decreases in
the course of bone loss whereas the volume and surface of the trabecular structure
changes in a different amount. This latter conclusion confirms former
findings that the shapes of the trabeculae change during bone loss (e.g.
from plate-like structure to rod-like structure) \cite{hildebrand1999}.

In this study we apply recently developed measures of complexity to 
characterise for the first time the temporal change of  
complex spatial structures of trabecular bone. We analyse the changes 
of the 3-dimensional micro-architecture
of distal tibia of patients who suffered on temporal immobilisation 
due to a disease or an accident on the left body side. Moreover, we 
investigate the effects of errors in positioning of volume of 
interest within the bone during the follow-up CT scans
and of choosing different segmentation parameters.

\section{Data and Methods}

For our analysis, we have used data from the
{\it Advanced Detection of Bone Quality (ADOQ)} study. 
In this framework, different subject groups 
were investigated in order to develop reference data for 
high-resolution 3D peripheral Quantitative Computer Tomography
(3D-pQCT). It includes 3D data from
patients suffering from ligament rupture or fracture of
the extremity, or from hemiplegic patients (half of the body
is paralysed). 

For each patient the 3D-pQCT investigations were performed
over a period of six months including baseline imaging and
four follow-up measurements with an interval of six weeks. The
measurements were performed at all extremities: left and
right distal radii and distal tibiae. In our investigation
we focused on distal tibia. We have selected four patients,
who have attended the follow-up measurements at least three times,
and who had suffered temporal immobility 
due to a disease or an accident on the left body side.

\begin{figure*}[bp]
  \centering
\includegraphics[width=.47\textwidth]{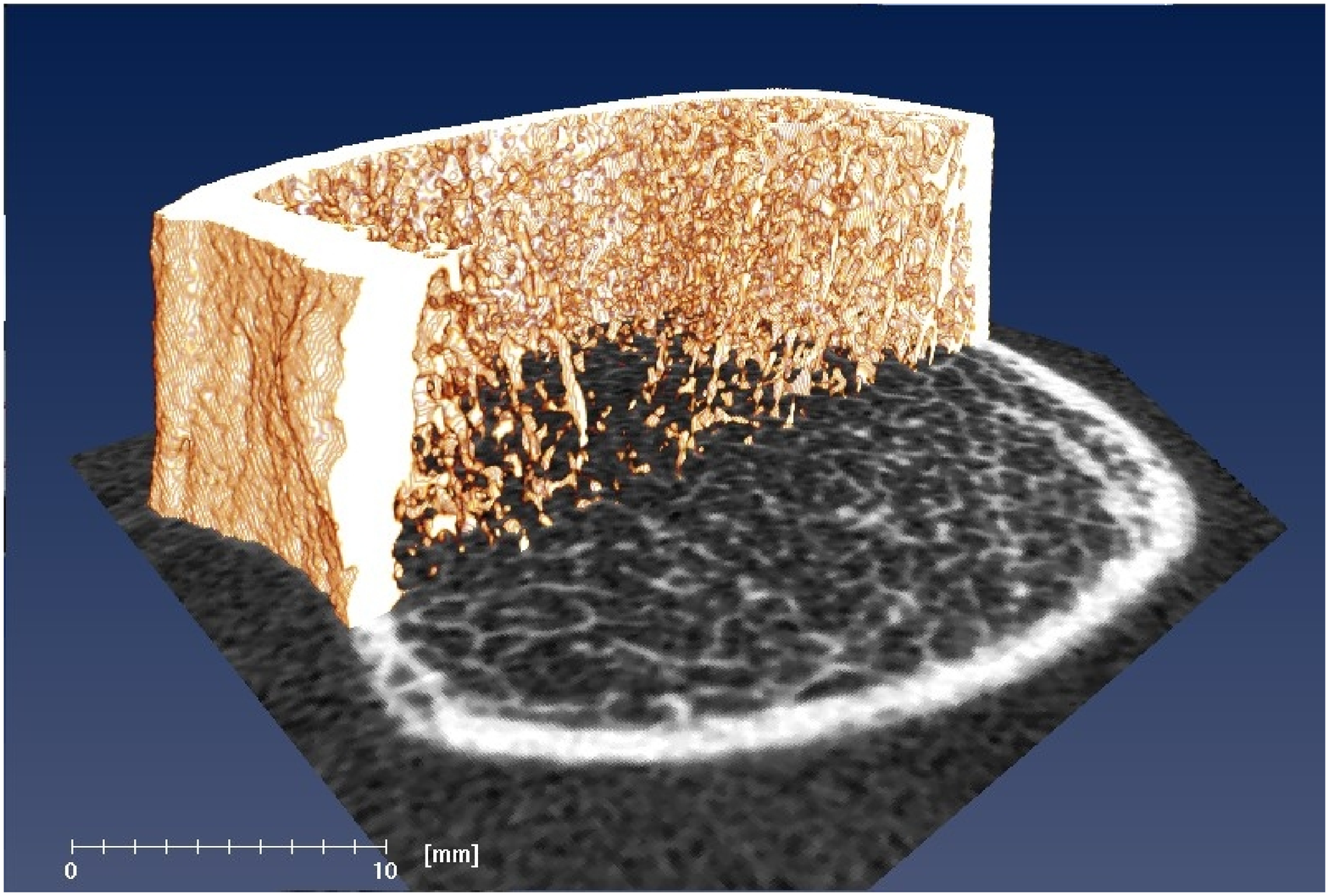}
\hspace*{4pt}
\includegraphics[width=.47\textwidth]{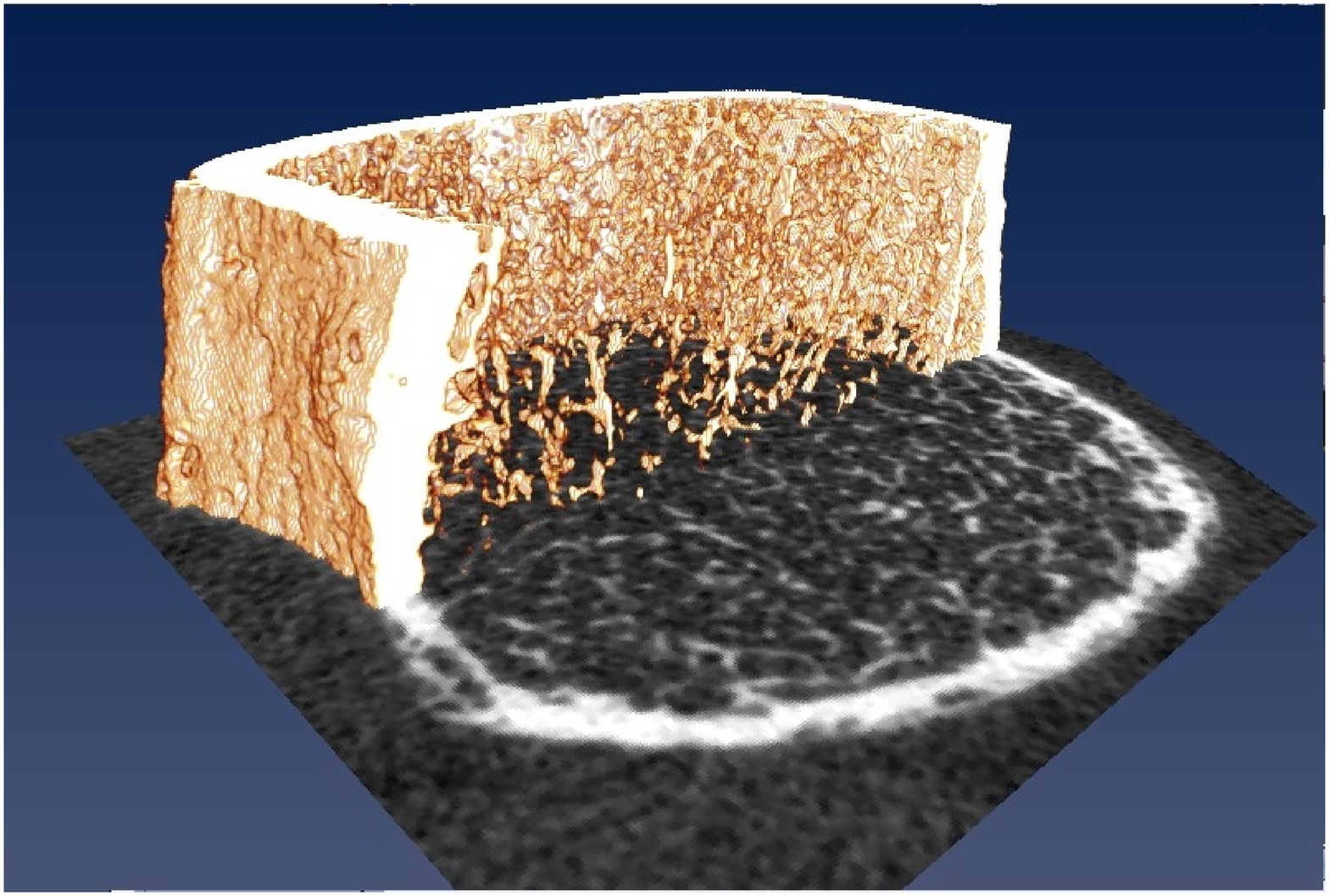}
\caption{3D-pQCT images of left distal tibia of patient 9
at the beginning of the study (left panel) and after four months 
at the end of the study (right panel) showing slight 
variations of the bone micro-architecture.
The lower, gray-scale plane represents
a 2D orthoslice, and the coloured volume rendering shows a
cutted part of the 3D image.
}\label{tibia}
\end{figure*}

The data were acquired in-vivo by XtremeCT
3D-pQCT scanner (Scanco Medical AG, Switzerland). The 3D-pQCT
images have an isotropic voxel size of 82~$\mu$m, the image matrix has
a size of 1536 voxels $\times$ 1536 voxels $\times$ 110~voxels (length of the
imaged region along the bone axis was 9.02~mm; Fig.~\ref{tibia}). 
The clinical situation of the
remaining patients is summarised in Tab.~\ref{diagnosis}.

\begin{table}[htb]
\tbl{Diagnosis of the four patients whos tibial bone structure
was imaged by 3D-pQCT and assessed by the structural measures of complexity.
\label{diagnosis}}{
\centering 
\begin{tabular}{p{1.cm}lp{1.5cm}}
\hline
Patient	&Clinical Diagnosis		&Follow-up \\
ID		&											&visits\\
\hline
5		&medial fracture on the left femoral neck	&5\\
9		&ligament rupture left knee					&4\\
11		&media infarct (stroke) with left-sided failure/ strong&\\
		&distal paralysis, caused wheel-chair dependence&3\\
14		&impression fracture on the left tibial plateau &4\\
\hline
\end{tabular}
}
\end{table}

The pQCT images were converted from 3D-pQCT data format into
a 3D mesh format for our analysis using the Amira software
platform (Mercury Computer Systems Inc.).

The comparison of follow-up 3D-pQCT scans was done by
considering the common regions overlapping in all scans (see
also Subsect.~\ref{local_mismatch}). 
This was necessary because the scan at
different time points causes slight displacements of the
imaged regions of distal tibia. To minimise deviations due to
dislocations/ displacements, the pQCT device establishes common
regions for all of the scans. However, small tilts cannot be
corrected with this method, hence leading to potential
variations in structural analysis.

For the quantification of the bone microarchitecture we focus
only on the trabecular part of the bone. Therefore, to
separate the trabecular bone from the rest of the imaged 3D
data we created manually volumes of interest (VOI). Then, in order to reduce
the level of noise in the pQCT data, the 3D images were filtered using
a 3D median filter. Bone and marrow voxels were separated by
using of a global thresholding; the threshold level was set
to 2800 units of the pQCT image (this value was found heuristically by looking at
the histogram of the attenuation values, but also see 
Subsect.~\ref{segmentation_threshold}). 

\begin{figure*}[tbp]
  \centering
\includegraphics[width=.47\textwidth]{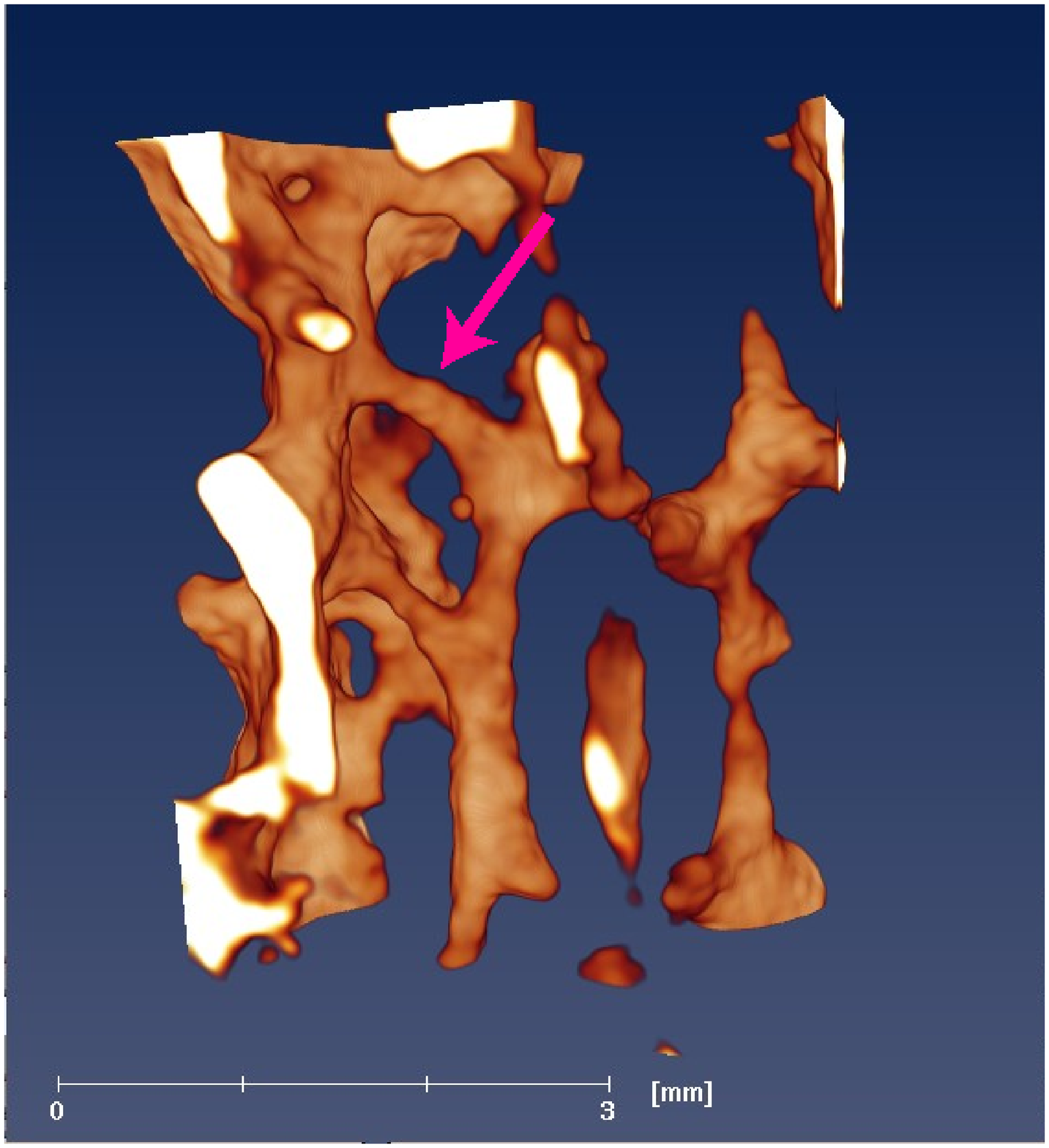}
\hspace*{7pt}
\includegraphics[width=.47\textwidth]{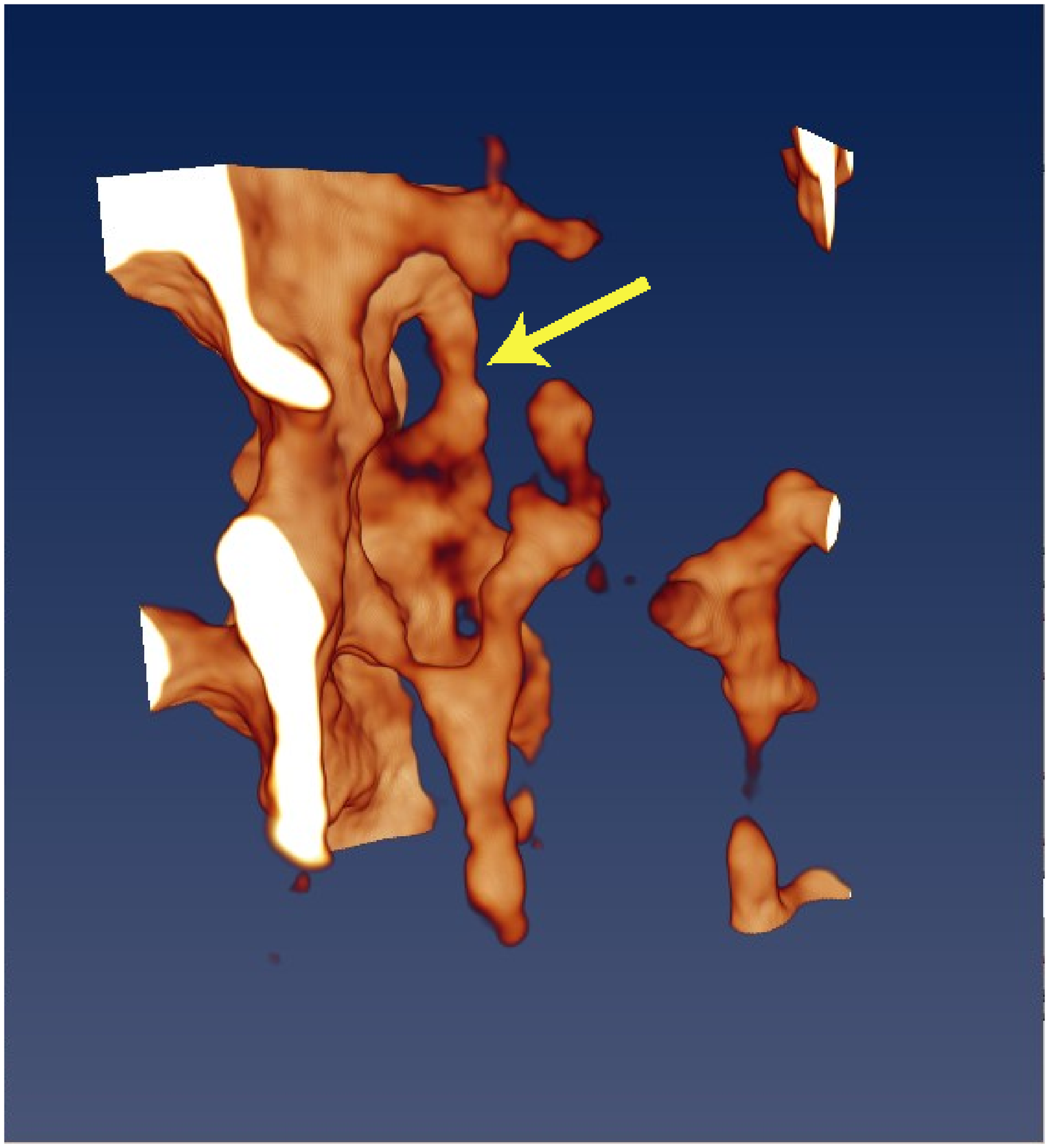}
\caption{A detail of the 3D-pQCT image shown in Fig.~\ref{tibia},
at the beginning of the study (left panel) and after four months 
at the end of the study (right panel).
The thinning and resorption of trabeculae is visible (magenta
arrow). But new trabeculae can also appear (yellow arrow).
}\label{trab}
\end{figure*}

At first, we calculated the classical measure bone volume fraction BV/TV which
provides  information about the amount of bone material.
Then we calculated recently introduced
structural measures of complexity (SMCs), which are basing on
symbol encoding, curvature, translational invariance and
local shape of trabecular bone elements (Tab.~\ref{smcs})
\cite{saparin2005,marwan2005bone,marwan2007epjst,marwan2009}.

\begin{table}[bp]
\tbl{Structural measures of complexity (SMCs) used in this
study.\label{smcs}}{
\centering 
\begin{tabular}{p{4.cm}p{5.cm}l}
\hline
\multirow{2}{4.cm}{\textit{SMCs based on symbol encoding} \cite{saparin1998,saparin2005}}
&3D Normalised Entropy of geometrical locations of bone tissue 	& $S_n$ \\
&Structure Complexity Index, based on bone volume fraction		&SCI$_{\text{BV/TV}}$\\
&3D Structure Complexity Index									&SCI$_{\text{3D}}$\\
&Surface Complexity Index										&SurfCI\\
&Surface Index of Global Ensemble 								&SurfIGE\\
\hline
\multirow{3}{4.cm}{\textit{Measures based on local shape and curvatures} \cite{marwan2005bone,esa2005,marwan2009}}
&Shape Complexity		&SHC\\
&Averaged Shape Index 	&ASHI\\
&Shape Mutual Information 	&MISH\\
&Marching Cubes Entropy Index 	&MCE\\
&Marching Cubes Complexity 		&MCC\\
&total curvature 				&$K$\\
&mean curvature 				&$H$\\
\hline
\multirow{2}{4.cm}{\textit{Measures assessing translational invariance} \cite{marwan2007epjst}}
&Lacunarity				&$\Lambda$\\
&Morans's $I$ index		&$I$\\
& 	&\\
\hline
\end{tabular}
}
\end{table}

SMCs based on symbol encoding use a limited
set of structural elements or symbols for the
quantification of the 3D structures.
The purpose of the symbol-encoding procedure is
to reduce the amount of information in a 3D-pQCT 
image to its essential structural composition by
representing the bone architecture. The modified 
symbol-encoding procedure for the
micro-CT data is based on an alphabet of three different
symbols: M, marrow voxel; I, internal bone
voxel; and S, surface bone voxel. Bone ``surface'' is a one
voxel-thick layer of bone voxels which are lying at
the boundary between two tissues: bone and marrow.
Based on distributions and joint distributions of the
symbols, the measures of complexity are defined. For example,
the surface complexity index SurfCI is defined by
\begin{equation}
SurfCI = - \sum_\text{VOI} p_{\text{surf}} \log_2 p_{\text{surf}}, 
\end{equation}
with $p_{\text{surf}}$ as the distribution of local surface indices
\begin{equation}
p_{\text{surf}} = \frac{p_{\text{loc}}(S)}{p_{\text{loc}}(I) + p_{\text{loc}}(S)}.
\end{equation}
This measure is calculated in a small moving box VOI.
SurfCI assesses the complexity of the distribution of bone
surface voxels in 3D. Other SMCs based on symbol encoding 
quantify the complexity or the distribution
of the trabecular structures \cite{saparin2005}.

The main idea behind the SMCs basing on the geometrical
shape is based on the fact that different 3D objects of the
same volume have different surfaces, depending on their
geometrical shape. For example, a long  cylinder (length is
much larger than radius) has a larger surface than a cube of
the same volume, whereas a sphere has the 
smallest possible surface for the same given volume.
Therefore, the ratio between volume and surface is used
in the definition of several SMCs, like the 
the local shape index
\begin{equation}\label{eq_shi} 
\sigma_{\text{loc}} =
\frac{S_{\text{bone}}}{S_{\text{sphere}}} \quad \text{with} \quad
S_{\text{sphere}} = 6 \sqrt[3]{ \pi \hat V_{\text{bone}}^2},
\end{equation} 
where $S_{\text{bone}}$ is the surface of trabecular bone
in a small box, $V_{\text{bone}}$ is its volume, and $S_{\text{sphere}}$
is the smallest possible surface for such volume (which corresponds
to the surface of a sphere). The average of the local shape index
ASHI, as well as the measures SHC and MISH, are able to 
distinguish between different shapes with the
same volume but whose surface differ, like plates and 
rods. MCE and MCC quantify the complexity of the bone surface \cite{marwan2009}.

Further measures base on fractal properties, translational invariance and 
small-scale spatial auto-correlation. For example, the lacunarity $\Lambda$
measures the variation of the bone fraction in small moving boxes by
using the ratio of the second and first moment of the distribution $n(s,r)$
of the bone density:
\begin{equation}\label{eq_lac}
\Lambda(r) = \frac{\mu_2(r)}{\mu_1^2(r)},
\end{equation}
where the first and second moments are
\begin{equation}
\mu_1(r)=\frac{1}{N}\sum_s s \, n(s,r)
\quad \text{and}
\quad \mu_2(r)=\frac{1}{N}\sum_s s^2 \, n(s,r),
\end{equation}
$N$ is the total number of boxes, $s$ the local bone
density and $r$ is the size of the moving box \cite{marwan2007epjst}. 
The lacunarity measures therefore how homogenously distributed
the trabecular bone is.

As we have mentioned, several SMCs are basing on analysis 
of ensembles of local properties estimated from a moving cubic box.
The size of the moving cubic box should be small enough to 
locally quantify structural elements, but large enough
to cover a sufficient surface necessary to distinguish between
different shapes and to reduce artefacts. Based on this 
requirements, we have found empirically a trade-off for the
box size to be $20 \times 20 \times 20$ voxels 
($1.64 \times 1.64 \times 1.64$~mm).

The SMCs have been calculated on the trabecular bone of the distal
tibia within the specified VOI. As the same bone region was
scanned at six weeks intervals, for the first time the
results allow us to study structural changes in the same
trabecular bone over time.

\section{Experiments to test stability and reproducibility of the 
structural measures}\label{stability_reproducibility}

Stability and reproducibility of the measures are crucial
requirements to any data analysis tools. In the follwoing we present
our studies of the influence of main
disturbing factors, including mismatch of VOIs and different
bone segmentation thresholds. In another study we have  
investigated effects of the image resolution \cite{marwan2009}.

\subsection{Local mismatch of VOI or common region}\label{local_mismatch}

For the study of structural changes during time, the
3D-pQCT images of the distal tibia have to be aquired at exactly
the same location in order to provide images of bone
appropriate for longitudinal evaluation. Therefore, the
software of the pQCT device calculates the common region between
consecutive images. Nevertheless, small deviations in the
positioning of the common region still remain, e.g.~due to tilts. Therefore, we
analysed the variation of the SMCs for an
artificial dislocation of the VOI.

In this numerical experiment we used 3D-pQCT data of the left distal tibia from
patient with ID 9. We reduced the original VOI by removing a
small 82~$\mu$m thick volume at its perimeter (that was defined
as 1~voxel from each side in the $xy$-plane) and applied small
random local displacements of the VOI in the $xy$-plane in the
interval of $\pm$0.95 voxels (corresponding $x$- and $y$-shifts were randomly
uniformly distributed between $-77$ and 77~$\mu$m). The
calculations were repeated 500 times in order to get
distributions of the measures from randomly displaced volume of interest.
(Tab.~\ref{variation_range}).

\begin{table}[hbtp]
\tbl{Median and variation range of the structural measures 
of complexity for small random local displacements of VOI in 
interval from $\pm$77~$\mu$m (corresponding approximately to less 
than one voxel). The results are obtained from 500 realisations 
of random displacements of the VOI in CT images of distal tibia 
of patient ID 9.\label{variation_range}}{
\centering 
\begin{tabular}{lrrrrr}
\hline
Measure	&Median	&Std.~Deviation	&$q_{0.05}$	&$q_{0.95}$	&Range (\%)\\
\hline

BV/TV	&0.22	&0.01	&0.20	&0.23	&13.14\\
SCI$_{\text{3D}}$	&0.76	&0.01	&0.73	&0.77	&4.39\\
SCI$_{\text{BV/TV}}$	&0.90	&0.01	&0.89	&0.91	&2.37\\
SurfCI	&0.86	&0.01	&0.84	&0.86	&2.22\\
SurfIGE	&0.70	&0.03	&0.69	&0.75	&9.39\\
$S_n$	&0.93	&0.00	&0.93	&0.93	&0.65\\
$H$	&1.03	&0.15	&0.91	&1.29	&37.01\\
$K$	&$-1.88$	&0.10	&$-2.08$	&$-1.75$	&17.59\\
Moran's $I$	&0.75	&0.01	&0.73	&0.75	&3.23\\
$\Lambda$	&0.41	&0.03	&0.35	&0.43	&19.36\\
ASHI	&1.73	&0.01	&1.73	&1.76	&1.70\\
MISH	&0.19	&0.00	&0.19	&0.19	&1.14\\
SHC	&0.22	&0.00	&0.22	&0.23	&2.99\\
MCC	&5.72	&0.03	&5.67	&5.75	&1.28\\
MCE	&0.65	&0.01	&0.64	&0.67	&4.83\\
\hline
\end{tabular}
}
\end{table}

Due to the random shifts by maximal of one voxel, the
variation range of the SMCs was between 0.65\%
($S_n$) and 37.0\% ($H$). As a variation range we consider here the
distance between the 5\%- and 95\%-quantiles normalised by the
median value of the measure.  Therefore, this performance
measure allows us to assess the influence of the local
displacement on the SMCs. We find that the measures BV/TV,
the curvatures $H$ and $K$ as well as $\Lambda$ are rather
sensitive to the displacement with variation ranges above
10\%. In contrast, the measures SCI$_{\text{BV/TV}}$, SurfCI, $S_n$, ASHI,
MISH, SHC and MCC are much less sensitive, below 3\%. The
remaining measures, SCI$_{\text{3D}}$, SurfIGE, Moran's $I$ and MCE are
slightly more sensitive with the variation range around 5\%.

These are crucial findings, because they imply that we cannot
interpret as significant relative variations in the range below 5\% (that is
a half of the maximal range of 10\% found in our simulations) in
such measures as BV/TV and SurfIGE, and even variations
around 10--20\% in the curvatures $H$, $K$ as well as 
$\Lambda$, as reflection of real structural changes, just
because it is impossible to exclude positioning errors due to the minor
shifts of the VOI during consecutive CT measurements. In the
interpretations of the results derived from the ADOQ data
in Sect.~\ref{ADOQ_study} we will take into account these error ranges.

\begin{figure*}[tbp]
  \centering
\includegraphics[width=\textwidth]{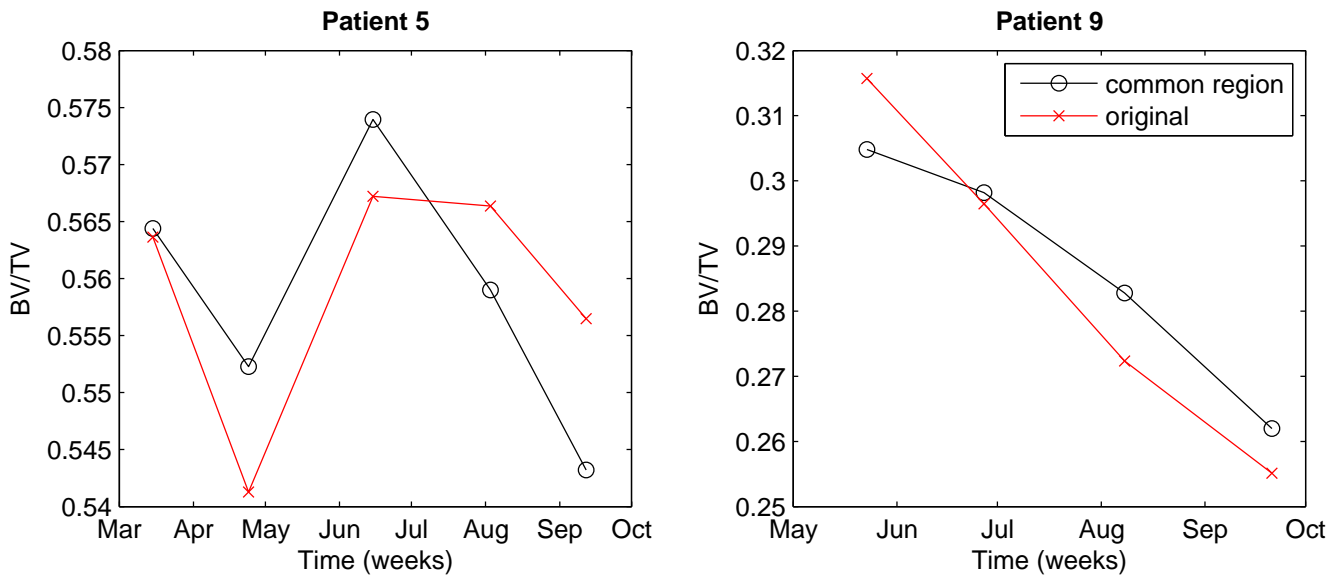}
\caption{The BV/TV calculated from the entire trabecular VOIs and the
common regions of left distal tibiae for patients 5 and 9.
}\label{bvtv_noshift}
\end{figure*}

\begin{figure*}[htbp]
  \centering
\includegraphics[width=\textwidth]{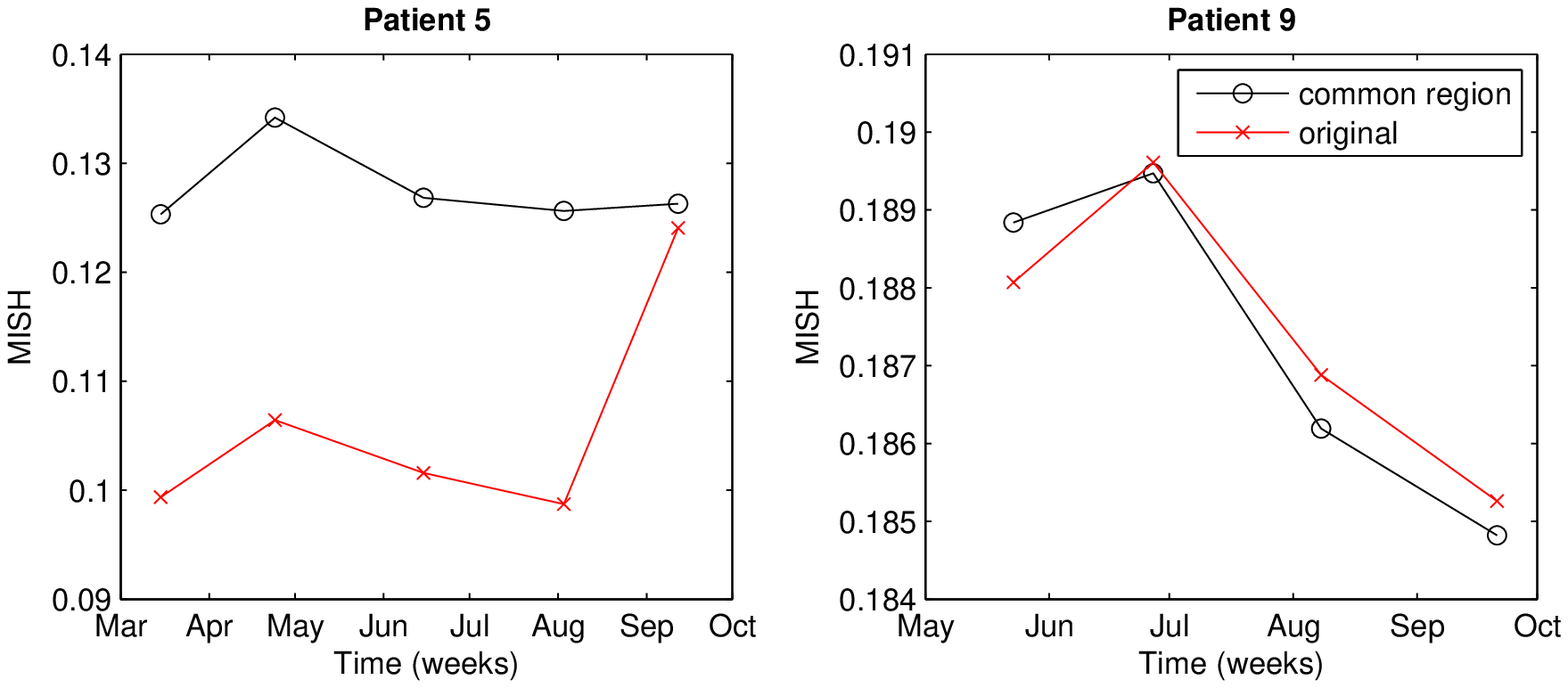}
\caption{The shape mutual information MISH calculated from the entire trabecular VOIs and the
common regions of left distal tibiae for patients 5 and 9.
}\label{mish_noshift}
\end{figure*}

Next, we compared the SMCs calculated from the
entire 3D-pQCT images and the measures obtained from the common
regions that match in different time-points. After evaluating the
structure in the common regions, we have repeated the
calculations of the SMCs for left distal tibia of patients
with ID 5 and 9 for the entire trabecular VOIs that are
larger than the common regions of the same measurements. The
differences in results derived from the common region and the entire
trabecular VOI are rather small for most of the
SMCs (Tab.~\ref{relative_differences}). In general, the typical trend in the
behaviour of the measures found in matched regions remains the
same when the entire VOIs are analysed (Fig.~\ref{bvtv_noshift}). Remarkable
differences in the changes in time occur in the measures SCI$_{\text{3D}}$ (only
patient 5), SurfIGE, MISH (only patient 5, see Fig.~\ref{mish_noshift}), MCE
and Moran's $I$ index. Large differences in the absolute values
were found in MISH, $H$ and $K$. However, for $H$ and $K$ the trend
remains very similar when matching regions and entire
trabecular VOI are compared.

\begin{table}[tp]
\tbl{Maximal and mean relative differences between the structural 
measures calculated for common region and for entire vertebral 
VOI of the images of patients ID 5 and 9.\label{relative_differences}}{
\centering 
\begin{tabular}{lrrrr}
\hline
\multirow{2}{1.5cm}{Measure}	&\multicolumn{2}{c}{Patient No.~5}&\multicolumn{2}{c}{Patient No.~9}\\
				&max.~dev.~(\%)	&mean dev.~(\%)	&max.~dev.~(\%)	&mean dev.~(\%)\\
\hline
BV/TV	&2.4	&1.4	&3.8	&2.6\\
SCI$_{\text{3D}}$	&1.5	&0.9	&0.9	&0.6\\
SCI$_{\text{BV/TV}}$	&0.5	&0.3	&0.6	&0.4\\
SurfCI	&1.1	&0.7	&1.0	&0.6\\
SurfIGE	&4.0	&2.3	&4.0	&2.5\\
$S_n$	&0.1	&0.0	&0.1	&0.0\\
$H$	&11.3	&7.1	&11.0	&6.5\\
$K$	&33.3	&21.3	&19.3	&13.1\\
Moran's $I$	&0.9	&0.5	&1.2	&0.8\\
$\Lambda$	&2.8	&1.4	&2.7	&1.7\\
ASHI	&1.1	&0.7	&0.4	&0.3\\
MISH	&21.8	&16.9	&0.4	&0.3\\
SHC	&6.3	&4.6	&0.8	&0.6\\
MCC	&0.8	&0.5	&0.3	&0.2\\
MCE	&2.3	&1.4	&2.1	&1.4\\
\hline
\end{tabular}
}\end{table}

\subsection{Dependence on the segmentation threshold}\label{segmentation_threshold}

The choice of the bone segmentation threshold is an important
step for evaluation of bone architecture. Therefore, we have studied how
the choice of segmentation threshold affects the different SMCs.
For this experiment we have used the 3D-pQCT
images from patient 5. The bone
segmentation threshold was set to values of 2300, 2800,
3200 and 4000 according to a threshold value
correspondingly set too low, optimal value, too high, and
extremely high.

\begin{figure*}[tbp]
  \centering
\includegraphics[width=\textwidth]{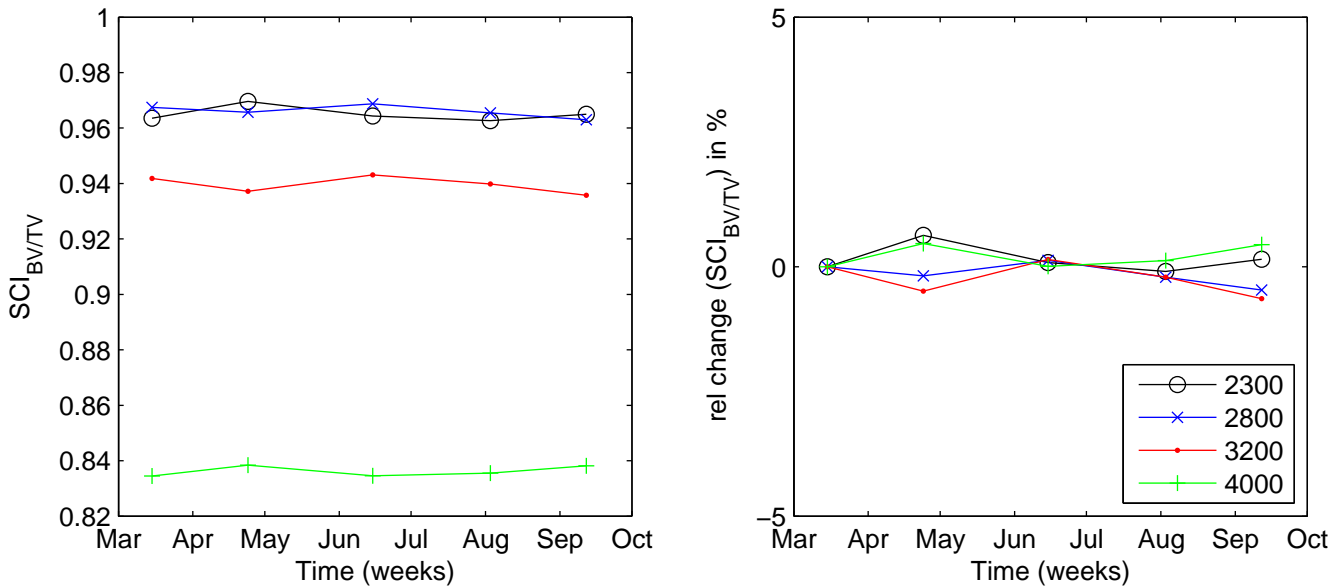}
\caption{Exemplary shown SMC for different values of bone segmentation
threshold reveals large difference in the absolute values for the
threshold level set extremely high (here 4000 pQCT units)
or low (2300 pQCT units).
}\label{sci_thresh}
\end{figure*}

\begin{figure*}[tbp]
  \centering
\includegraphics[width=\textwidth]{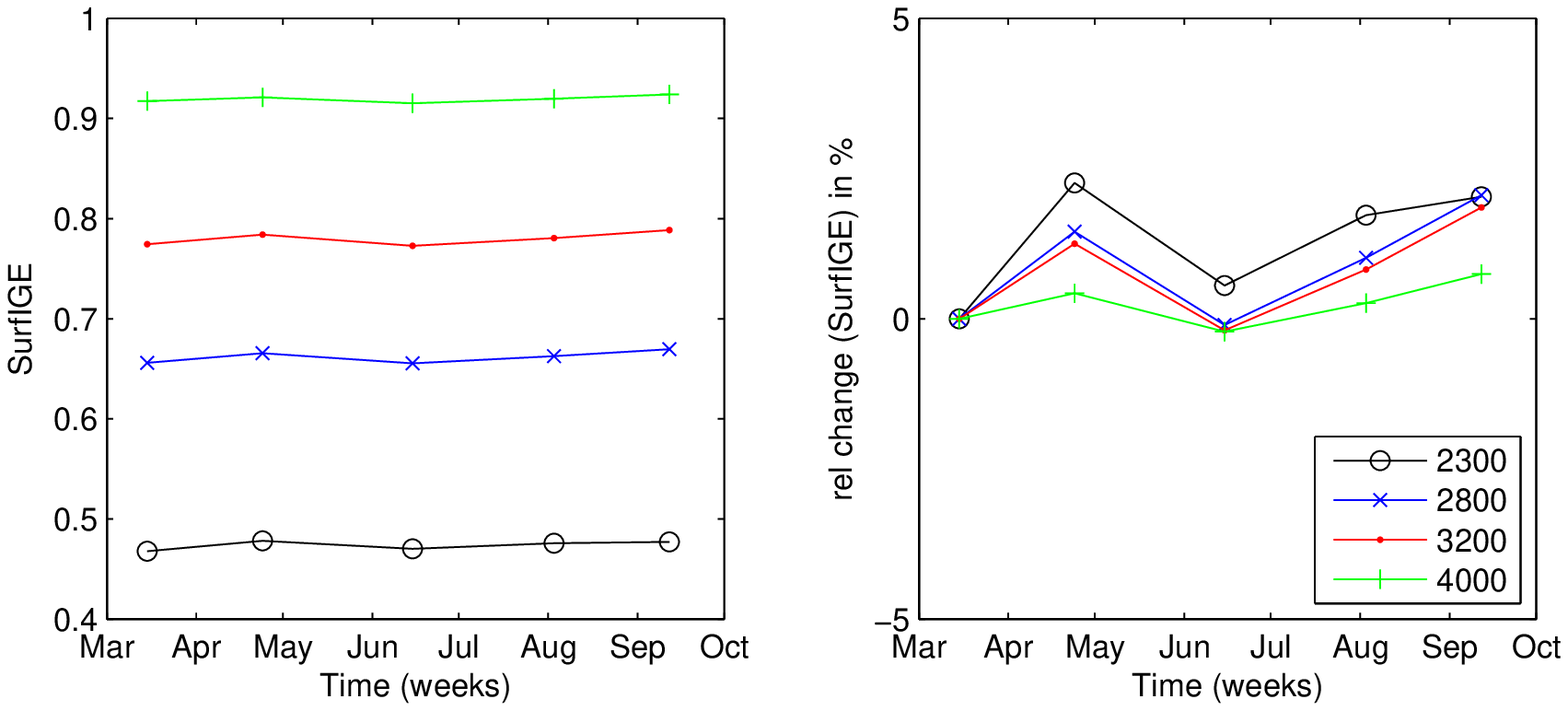}
\caption{Exemplary shown SMC for different values of bone segmentation
threshold reveals different absolute values but same qualtitative
behaviour.
}\label{surfige_thresh}
\end{figure*}

We found that absolute values of all SMCs depend on the segmentation
thresholds. For some measures like SCI$_{\text{3D}}$, SCI$_{\text{BV/TV}}$,
$K$ and $\Lambda$, their absolute values for the extreme
thresholds 2300 and 4000 differ much from the values of the
thresholds 2800 and 3200 (Fig.~\ref{sci_thresh}). For several measures, as
BV/TV, SCI$_{\text{3D}}$, SCI$_{\text{BV/TV}}$, SurfCI, $S_n$, $H$, $K$, 
$\Lambda$, MISH and
MCC we found a different trend only for the extreme low and
extreme high threshold levels, whereas for levels of 2800 and
3200, the trend remained similar. Similar evolution in time were found
for the other measures SurfIGE, Moran's $I$, ASHI, SHC and MCE
indicating that the trend (or qualitative behaviour) of these
measures is not as sensitive to the choice of the threshold as
of the other measures (Fig.~\ref{surfige_thresh}). Large differences for the
longitudinal behaviour, and hence high sensitivity, were found for the measures
$H$, $K$ and MCC.

These results confirm that our choice of the bone segmentation
threshold at the level of 2800 is appropriate,
because more extreme threshold values like 2300 or 4000 reveal
more different results (but in the same direction). Moreover,
this numerical experiment demonstrates that a slight change of 
the threshold would not affect the qualitative findings
how the SMCs evolve with time.

\section{Evaluation of 3D-pQCT data from ADOQ Study}\label{ADOQ_study}

\begin{figure*}[tbp]
  \centering
\includegraphics[width=\textwidth]{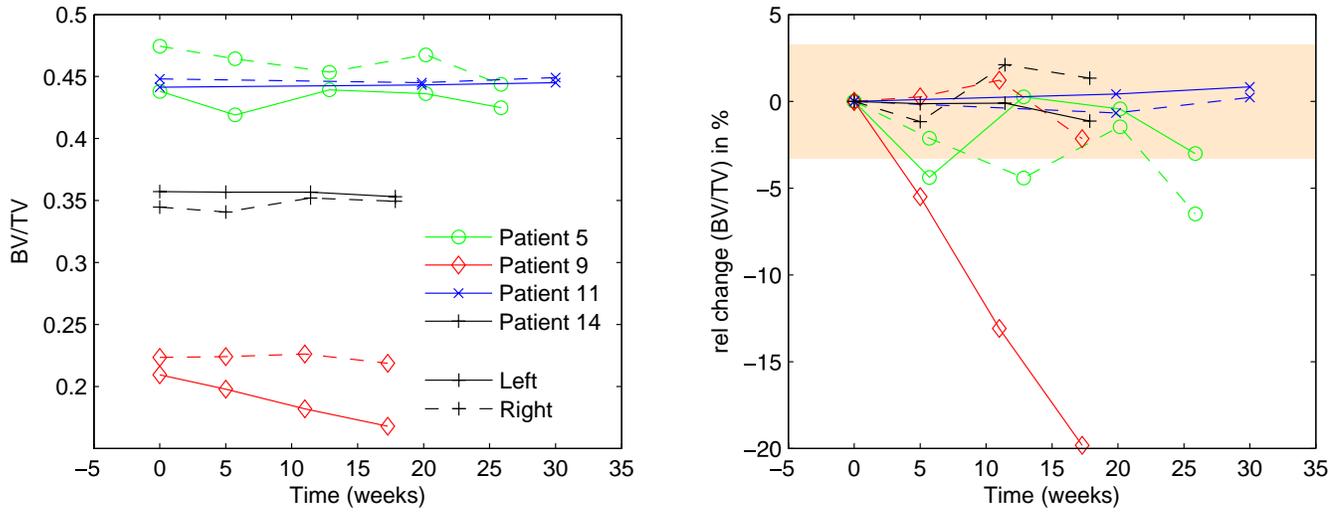}
\caption{Bone volume fraction BV/TV expressed in \% (left panel) and 
its relative change (right panel) during the consecutive measurements 
after trauma or disease caused immobility of the left limb. 
The rose-shaded region marks the range of variation caused by positioning errors.
}\label{res_1}
\end{figure*}

\begin{figure*}[tbp]
  \centering
\includegraphics[width=\textwidth]{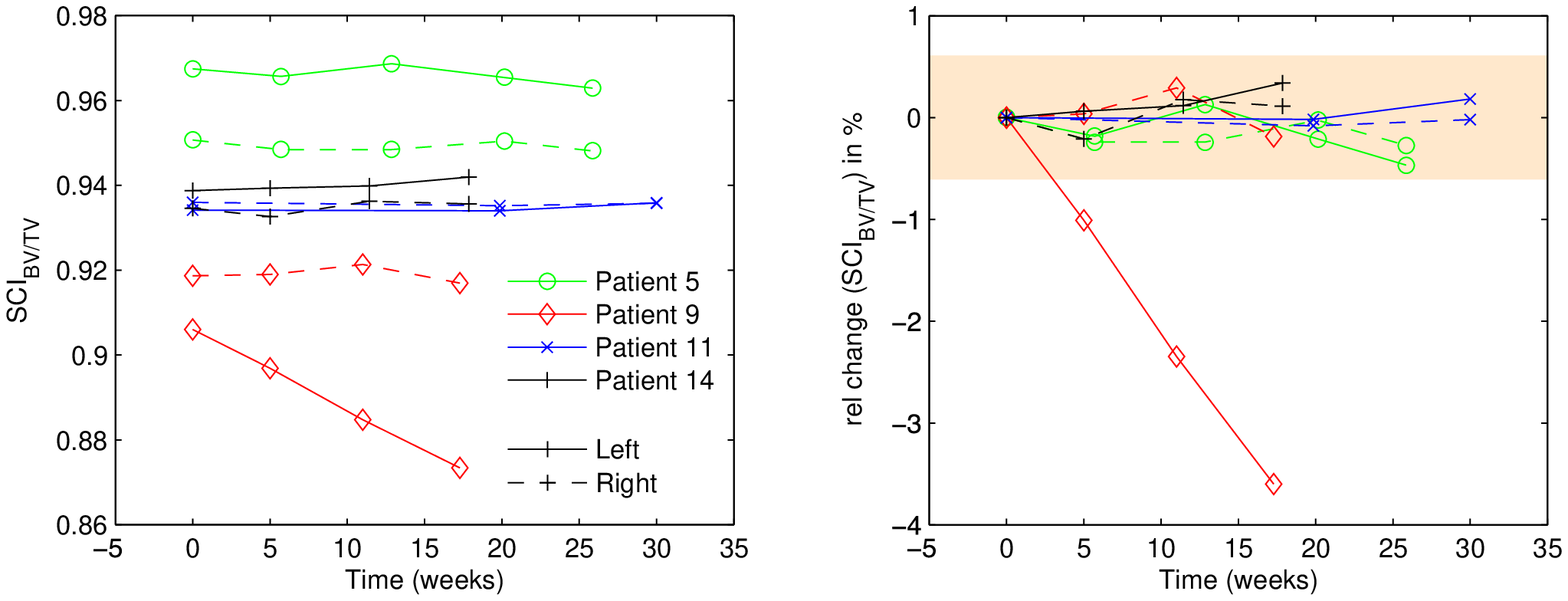}
\caption{Structural complexity measured by SCI$_{\text{BV/TV}}$ (left panel) 
and its relative change (right panel) during the consecutive measurements after 
trauma/ disease caused immobility of the left limb. The rose-shaded region 
marks the range of variation caused by positioning errors.
}\label{res_3}
\end{figure*}

\begin{figure*}[tbp]
  \centering
\includegraphics[width=\textwidth]{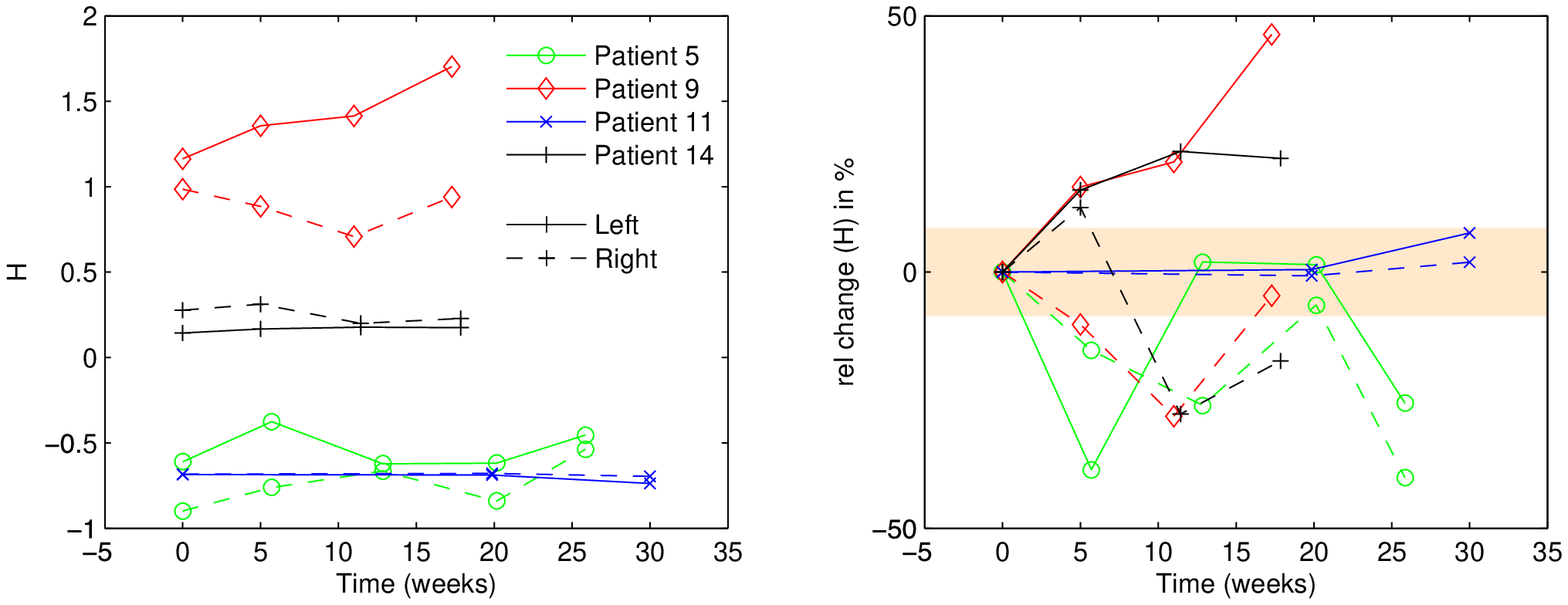}
\caption{Mean curvature $H$ (in mm$^{-1}$, left panel) and its relative 
change (right panel) during the consecutive measurements after 
trauma/ disease caused immobilty of the left limb. The rose-shaded 
region marks the range of variation caused by positioning errors.
}\label{res_7}
\end{figure*}

\begin{figure*}[tbp]
  \centering
\includegraphics[width=\textwidth]{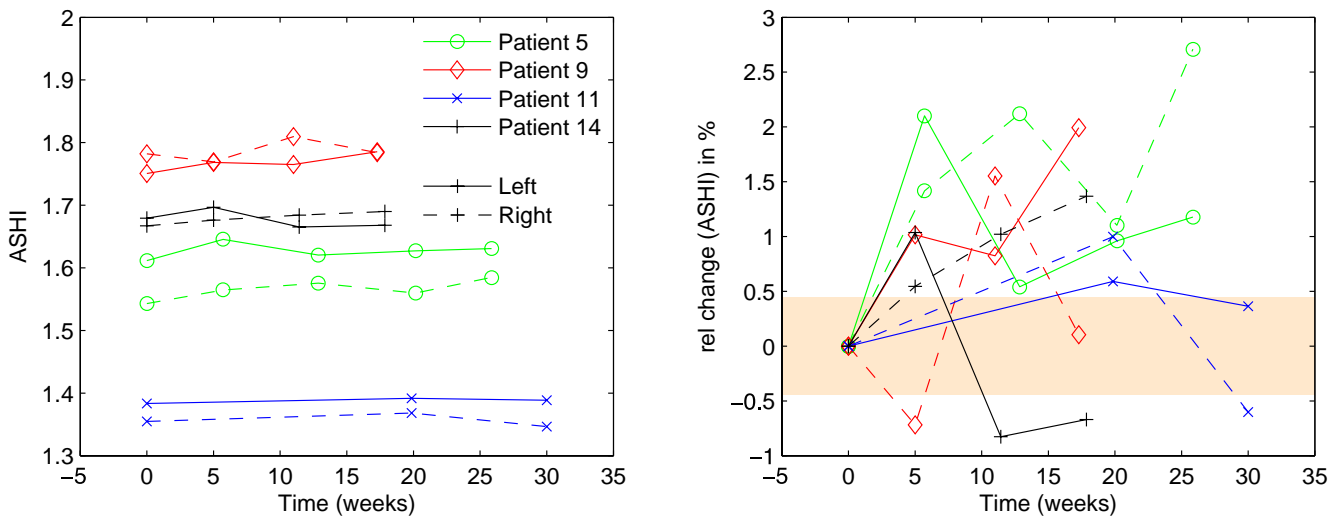}
\caption{Shape complexity measured by ASHI (left panel) and its 
relative change (right panel) during the consecutive measurements 
after trauma/ disease caused immobility of the left limb.
}\label{res_11}
\end{figure*}

\begin{figure*}[tbp]
  \centering
\includegraphics[width=\textwidth]{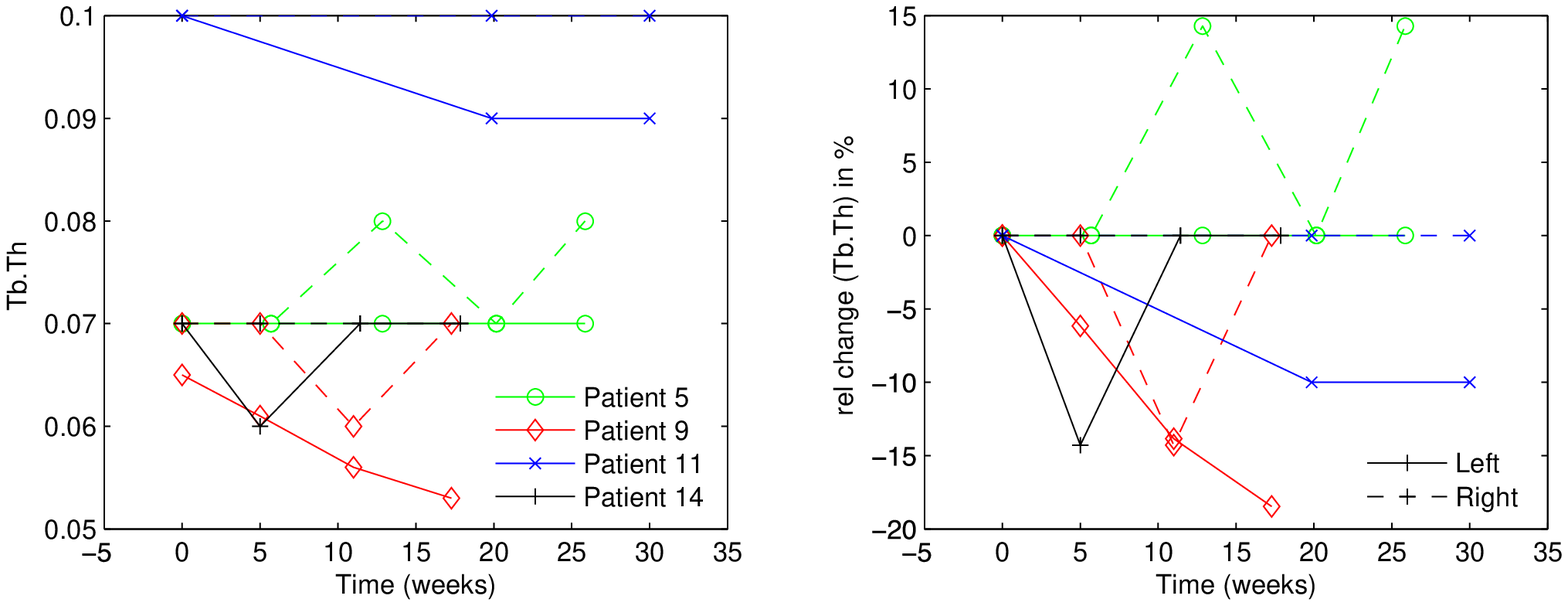}
\caption{Trabecular thickness Tb.Th (in mm, left panel) and its 
relative change (right panel) during the consecutive measurements 
after trauma/ disease caused immobility of the left limb.
}\label{res_18}
\end{figure*}

The set of 3D structural measures of complexity was
applied to evaluate the architectural changes of 
trabecular bone in patients
of temporal immobility, imaged by 3D-pQCT  
obtained during the ADOQ study. In order to compare SMCs with
classical histomorphometry we use the trabecular thickness Tb.Th.
The findings in regards of the
stability and reproducibility of the structural measures
reported in Section~\ref{stability_reproducibility} were used to validate the range if
changes of the SMCs have been found in longitudinal 
study (except the Tb.Th, because this
measure was calculated directly by the pQCT device). The
results of the evaluation are shown in Figs.~\ref{res_1}--\ref{res_18}.

Considering the BV/TV of all the four analysed patients and
comparing the left and right distal tibiae, we observe bone
degradation in patients 5 and 9, whereas the BV/TV for
patients 11 and 14 remained almost constant (Fig.~\ref{res_1}). This
is especially surprising for patient 11, because after the
stroke he was partly disabled and needed a wheel-chair. Most
likely the rehabilitation therapy after the stroke was very
effective. In contrast, patient 9 experienced a clear
decrease in BV/TV, in particular on the left distal tibia,
probably because the patient saved his left leg. The BV/TV of
the patient 5 tibiae decreased slightly.

For {\bf patient 11} (left-sided paralysis after stroke) we found
that all measures did not change significantly during the
investigated interval (see 
Figs.~\ref{res_1}--\ref{res_18}). Except for ASHI and MISH, whose range
of change is rather small, all relative changes remained
within the range of measurement-to-measurement variability
due to positioning of the VOI. Moreover, from our analysis we
can conclude that this patient (as well as patient 5) has
probably the strongest bone among all considered patients:
BV/TV of patients 11 and 5 has the highest values, its
trabecular architecture is very complex, as it is indicated by
high SCI$_{\text{3D}}$, the mean curvature $H$ is lowest (and even
negative), ASHI has lowest values suggesting more convex
trabecular structure, what is typical for a dense trabecular
bone. From the viewpoint of bone morphometry, the proximal
tibiae of patient 11 are also characterised by the
highest Tb.Th (Figs.~\ref{res_18}).

{\bf Patient 14} (impression fracture left tibial plateau) has an
intermediate level of bone volume fraction and complexity of
trabecular bone structure (Figs.~\ref{res_1}--\ref{res_18}). For this
patient most of the measures have their values between the
extreme values of patients 9 and 11. However, during the
investigated period of time, the trabecular bone has not
changed much as it is indicated by the structural measures
which vary only within their error range. We found a slight
increase of the surface complexity of the affected left
distal tibial trabecular bone after week 10, as revealed by
the measures SurfCI, $S_n$, ASHI and SHC (all $<$ 1\%). Moreover,
the mean curvature $H$ increased slightly, suggesting a small
amount of thinning of the trabeculae.

The SMCs for {\bf patient 5} (fracture on the left femoral neck)
reveal that this patient has also strong bone, as observed by
high bone density, high complexity of the trabecular
architecture, and low values of the mean curvature $H$ (that is
also negative). During the investigation
period, the trabecular bone composition is only slightly changing 
(Figs.~\ref{res_1}--\ref{res_18}). The bone density decreased up to $-$5\%, the
structural complexity as quantified by SCI$_{\text{3D}}$ decreased only
by $-$2\%, and the variability of local trabecular shapes
decreased (measured by an increase of ASHI by 2.5\%, MISH by
15\%, and a decrease of SHC by $-$7\%). $H$
varied during the investigated period, but finally increased
(25\%). The other measures vary only slightly or do not allow
reliable interpretations because the range of variations is
within the measurement-to-measurement error range.
Nevertheless, it is remarkable that the left and right side
seems to be affected in a similar way; the structural
measures vary on both sides with same amplitudes. However,
for several measures we found anti-correlation between the
values obtained from the left and right side (e.g.~in BV/TV,
SCI$_{\text{3D}}$, total curvature $K$, ASHI or MCE). This could be caused
by an alternative loading of the left and the right sides
(starting with the right because the left side suffered the
fracture).

In contrast, {\bf patient 9} (ligament rupture) turned out to be
the patient with lowest bone density and lowest complexity of
the trabecular microarchitecture. This patient experienced a
dramatic degradation of trabecular bone as expressed by a decrease of
bone volume fraction (up to $-$20\%) and significant changes of the
trabecular structure during the investigation period of 17
weeks, in particular on the left side (which was the affected
side by the ligament rupture; see Figs.~\ref{res_1}--\ref{res_18}). The
thinning of the trabeculae is well expressed by the increase
of the mean curvature $H$ (50\%) and the decrease of Tb.Th ($-$20\%).
The structural complexity as quantified by SCI$_{\text{3D}}$ decreased by
$-$7\% (this change is also confirmed by the other SMCs like
SCI$_{\text{BV/TV}}$, SurfCI, and SurfIGE). The degradation of the
trabecular structure caused an increase of the variability of
the local shape of bone's structural elements, expressed by
a decreasing of MISH by $-$4\%.

\section{Conclusions}

We have applied recently introduced structural measures of complexity (SMC), 
which are based on symbol encoding, curvature, translational invariance and
local shape of trabecular bone elements, to
trabecular bone structure in distal tibia. For the first time,
we have been able to study temporal alterations in 3D trabecular
bone structure using SMCs.

The developed SMCs are able to detect and assess the temporal 
alterations in the trabecular bone structure in distal tibia
due to temporal immobility. The detected structural variations 
have been different from subject to subject due to different
disease or sites of the fractures, as well as different treatment. 
For two patients the range of
change of some measures was smaller than their
measurement-to-measurement variability range due to the
errors in positioning of the imaged part of the tibia, suggesting
a positive effect of the preventive treatment. 
The developed SMCs based on the shape index are sensitive to
structural changes; their range of change is significant and
exceeds the interval of measurement-to-measurement variation
due to different positioning of the analysed limb.

Moreover, we have investigated the effect of small displacements
of the imaged volume during the pQCT scan and influence of different bone
segmentation thresholds. Using simulations of artifical position 
errors of the volume of interest we have been able to define a 95\%
variation (confidence) level for the SMCs. 
In the future, the SMCs will be further systematically
studied by applying models for bone remodelling 
(e.g.~\cite{huiskes2000,weinkamer2004,rusconi2008}).

Our findings support that the recently proposed SMCs are useful
to investigate trabecular bone alterations
\cite{saparin2005,marwan2005bone,marwan2007epjst,marwan2009}. This
study has demonstrated their abilities to quantify longitudinal 
changes. In particular they can be applied in clinical 
diagnosis or monitoring when 3D-pQCT imaging is available.

\section*{Acknowledgments}
\noindent This study was co-funded by
the Microgravity Application Program/ Biotechnology from the 
Human Spaceflight Program of the European Space Agency 
(project MAP AO-2004-125), by the European Commission ADOQ study
(Contract QLK-CT-2002-02363, Key action No.~6) and the Swiss government,
and received support from Siemens AG and Scanco Medical AG. 
We thank Hans-Chrisian Hege and 
Steffen Prohaska, Zuse Institute Berlin (ZIB), for the support
and help with the Amira software, as well as Jesper Skovhus Thomsen,
Peter Fratzel and Wolfgang Gowin
for fruitful discussions.

\bibliographystyle{ws-ijbc}
\bibliography{mybibs,bone}

\end{document}